\documentclass[3p,times,twocolumn]{elsarticle}

\usepackage{ecrc}

\volume{00}

\firstpage{1}

\journalname{Nuclear Physics B Proceedings Supplement}

\runauth{D.~Boito et al.}

\jid{nuphbp}

\jnltitlelogo{Nuclear Physics B Proceedings Supplement}

\usepackage{graphicx}

\begin{document}

\begin{frontmatter}

\title{The strong coupling from tau decays without prejudice}
\author{Diogo Boito$^a$, Maarten Golterman$^{b,c,}$\fnref{label1}}
\fntext[label1]{Speaker at workshop}
\author{
Matthias Jamin$^d$, Andisheh Mahdavi$^c$,
Kim Maltman$^e$, James Osborne$^{c,f}$, Santiago Peris$^g$}
\address{$^a$Physik Department T31, Technische Universit\"at M\"unchen,
James-Franck-Stra\ss e 1, D-85748 Garching, Germany}
\address{$^b$Institut de F\'\i sica d'Altes Energies, Universitat Aut\`onoma de Barcelona, E-08193 Bellaterra, Barcelona, Spain}
\address{$^c$Department of Physics and Astronomy, 
San Francisco State University, San Francisco, CA 94132, USA}
\address{$^d$Instituci\'o Catalana de Recerca i Estudis Avan\c cats (ICREA), 
IFAE,  Universitat Aut\`onoma de Barcelona, E-08193 Bellaterra,
Barcelona, Spain}
\address{$^e$Department of Mathematics and Statistics,
York University,  Toronto, ON Canada M3J~1P3 and
CSSM, University of Adelaide, Adelaide, SA~5005 Australia}
\address{$^f$Department of Physics, University of Wisconsin, Madison,
WI 53706, USA}
\address{$^g$Department of Physics, Universitat Aut\`onoma de Barcelona, E-08193 Bellaterra, Barcelona, Spain}

\begin{abstract}
We review our recent determination of the strong coupling $\alpha_s$
from the OPAL
data for non-strange hadronic tau decays.   We find that $\alpha_s(m^2_\tau)
=0.325\pm 0.018$ using fixed-order perturbation theory, and
$\alpha_s(m^2_\tau)=0.347\pm 0.025$ using contour-improved perturbation
theory.   At present, these values supersede any  earlier determinations
of the strong coupling from hadronic tau decays, including those from
ALEPH data.
\end{abstract}




\end{frontmatter}

\setcounter{footnote}{0}

\begin{figure}[t]
\includegraphics[width=.48\textwidth]{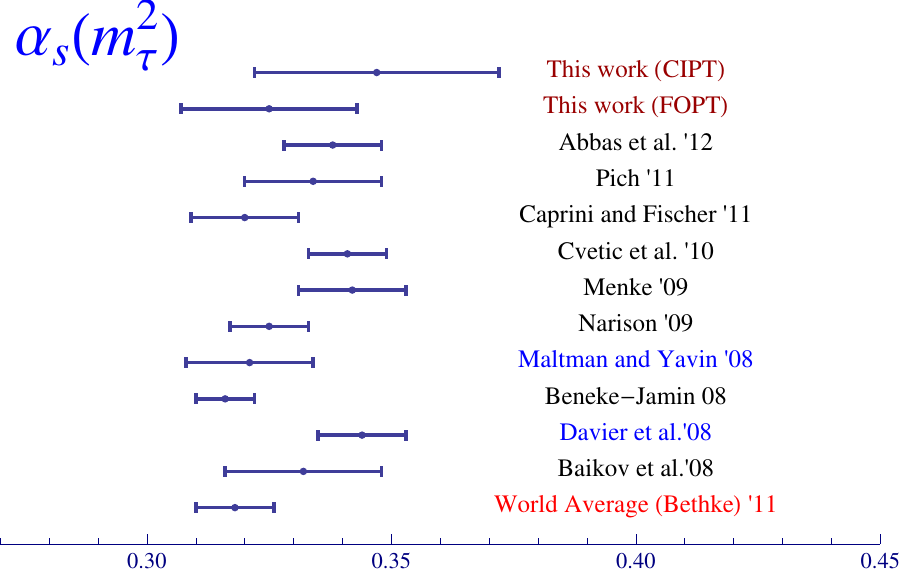}
\caption{Recent values for $\alpha_s(m^2_\tau)$ from hadronic tau decays.
For references, see Ref.~\cite{Bethke}.}
\label{alpha_s_after}
\end{figure}

{\bf 1.}
Figure~\ref{alpha_s_after} shows a number of recent determinations of the strong
coupling, $\alpha_s(m^2_\tau)$, from hadronic tau decays.  The two values at the top are recent determinations \cite{Betal2} from OPAL data \cite{OPAL}, and have signficantly larger errors than all the other determinations shown.
The reason for these larger errors is twofold:  (1)  errors in the non-perturbative part of the sum rules used in order to extract $\alpha_s(m^2_\tau)$ have been
systematically underestimated in earlier works;  (2)  all other values shown in Fig.~\ref{alpha_s_after} used ALEPH data, with correlations in which the effects of unfolding the spectrum have inadvertently been omitted \cite{Betal10}.
Here we give a brief overview of our analysis \cite{Betal2,Betal1}, comparing it with the standard approach used in for instance Ref.~\cite{Detal08}.  We note
that for the values shown in Fig.~\ref{alpha_s_after} only those of Refs.~\cite{Betal2,Detal08,MY08} are based directly on data; all others used estimates for the non-perturbative part from Ref.~\cite{Detal08}.

{\bf 2.}
All tau-based determinations of $\alpha_s(m^2_\tau)$ start from the sum rule
\begin{eqnarray}
\label{sumrule}
\int_{0}^{s_0}\!\!\! ds\;w(s)\;\rho_{\rm exp}(s)&\!=\!&-\frac{1}{2\pi i}\oint_{|z|=s_0}
\!\!\!\!\! dz\;w(z)\;\Pi_{\rm OPE}(z)\nonumber\\
&&\hspace{-1cm}-\frac{1}{\pi}\int_{s_0}^\infty\!\!\! ds\;w(s)\;{\rm Im}\;\Pi_{\rm DV}(s)\ ,
\end{eqnarray}
in which $w(s)$ is a polynomial weight, $\rho_{\rm exp}(s)$ is the inclusive,
non-strange spectral function taken from experiment, $\Pi_{\rm OPE}(z)$ is
the operator product expansion (OPE) expression for the non-strange flavor off-diagonal vacuum polarization with perturbation theory constituting the (dimension $D=0$) dominant part,
and
\begin{equation}
\label{DV}
\Pi_{\rm DV}(z)\equiv\Pi_{\rm QCD}(z)-\Pi_{\rm OPE}(z)
\end{equation}
the correction for using $\Pi_{\rm OPE}(z)$ instead of the (unknown) exact
$\Pi_{\rm QCD}(z)$.  It is important to include an estimate for the duality-violating
(DV) term in Eq.~(\ref{sumrule}), because duality violations are not small
near the Minkowski axis in the complex $z$ plane, from which the integral
on the left-hand side of Eq.~(\ref{sumrule}) originates.   In more physical
terms, 
the OPE does not capture the hadronic resonances which are
clearly visible in the experimental spectrum $\rho_{\rm exp}(s)$.

While the perturbative part of $\Pi_{\rm OPE}(z)$ is known to 4th order
in $\alpha_s(m^2_\tau)$ \cite{BCK} and the condensate corrections are 
expected to give a good estimate for non-perturbative corrections for
$|z|=s_0$ large and {\it away} from the Minkowski axis, we need to use a
model in order to estimate the effects of the DV term in Eq.~(\ref{sumrule}).
Based on an extensive study of models for the resonances seen in
$\rho_{\rm exp}(s)$, we have used the {\it ansatz} (see Ref.~\cite{Betal1}
and references therein)
\begin{equation}
\label{ansatz}
\frac{1}{\pi}\;{\rm Im}\;\Pi_{\rm DV}(s)=e^{-\gamma s-\delta}
\sin{(\alpha+\beta s)}\ .
\end{equation}

{\bf 3.} 
This very brief overview of the theory (for more details and references, see
Refs.~\cite{Betal1,Betal2}) makes it possible to compare the ``standard''
analysis of Refs.~\cite{OPAL,Detal08}) with ours.  In the standard analysis:
\begin{itemize}
\item{} It is assumed that duality violations can be neglected if one uses
``pinched'' weights $w(s)$, which always include factors $(1-s/s_0)^n$, with $n=2$ or $n=3$.  Such weights are thus necessarily polynomials of at least
degree 2.
\item{} One chooses $s_0=m_\tau^2$, and 5 pinched weights of degrees 3 to 7,
generating 5 data points, to which one fits 4 parameters, $\alpha_s(m^2_\tau)$ and the dimension 4, 6 and 8 OPE condensates.
We note that three of the five employed moments have problematic perturbative
behavior \cite{BBJ}.
\item{} This assumes that OPE condensates of dimension 10 through 16
vanish, because a term of order $s^k$ in the weight $w(s)$ picks out a
term of dimension $D=2k+2$ in the OPE.   This assumption was shown to be 
not self-consistent in Ref.~\cite{MY08}.
\end{itemize}
In contrast, in our work:
\begin{itemize}
\item{}  Our main fit uses the data for the vector channel, and the weight $w=1$, for which no OPE terms beyond
perturbation theory contribute.\footnote{Up to numerically negligible $\alpha_s$ corrections to these terms \cite{Betal1}.}
\item{}   We let $s_0$ vary over an interval $[s_{min},m_\tau^2]$, with
$s_{min}$ determined by the quality and stability of the fit; typically,
$s_{min}\approx 1.5$~GeV$^2$.   This provides more (correlated)
data compared to choosing $s_0=m_\tau^2$.
\item{}  We include duality violations. Our main fit thus has 5 parameters, $\alpha_s(m^2_\tau)$ and the 4 parameters of Eq.~(\ref{ansatz}) for the vector channel.
\item{}  We check consistency using also the axial-channel data, and
adding other weights up to degree 3 (always retaining all required terms
in the OPE, {\it i.e.}, to dimension 8.)
\end{itemize}
For more discussion of our strategy, we refer to Ref.~\cite{Betal1}.

\begin{figure}[t]
\includegraphics[width=.48\textwidth]{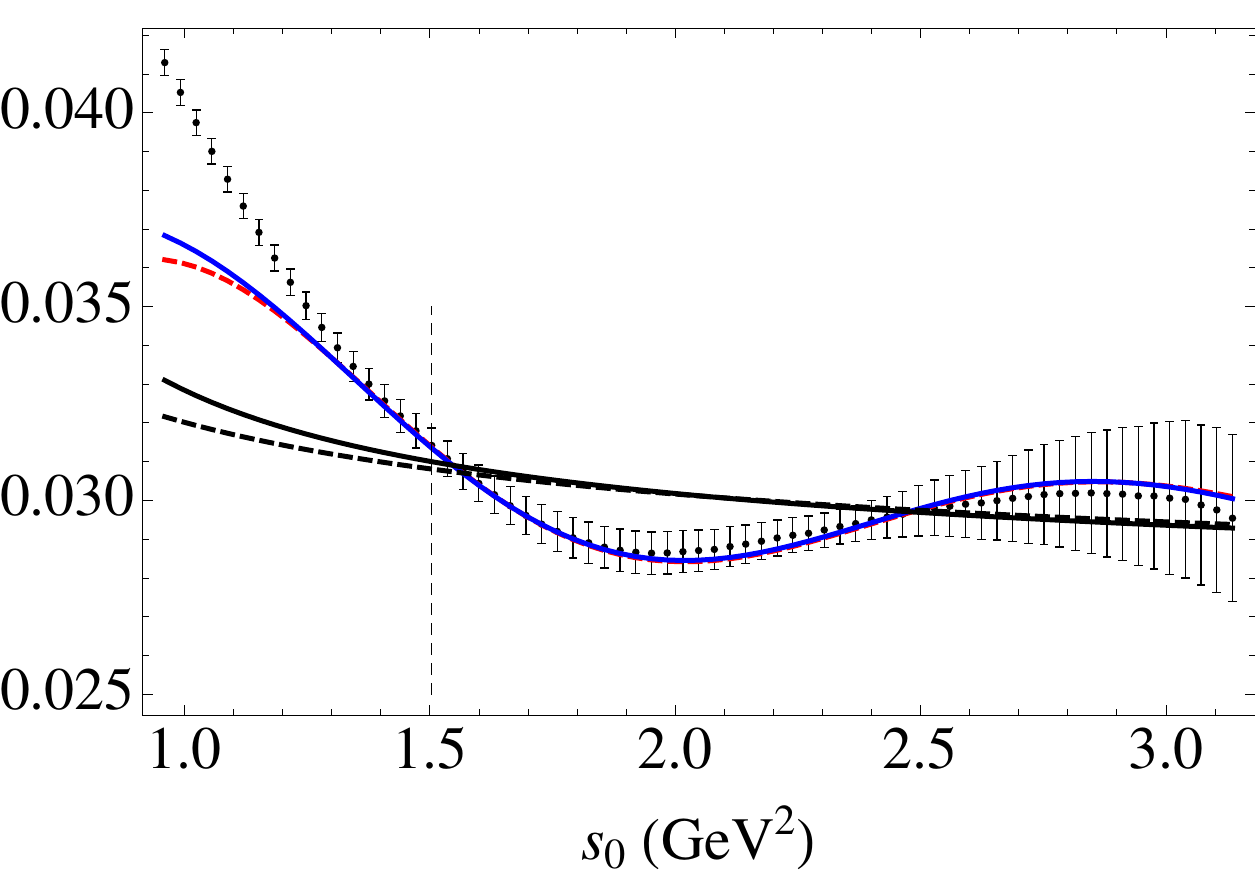}
\caption{Vector channel fit of Eq.~(\ref{sumrule}) with $w=1$ using OPAL
data. CIPT fits are shown in red (dashed) and
FOPT in blue (solid).
The (much flatter) black curves represent the OPE parts of the fits.  The vertical dashed line
indicates the location of $s_{min}$.}
\label{VVw1}
\end{figure}

\begin{figure}[t]
\includegraphics[width=.48\textwidth]{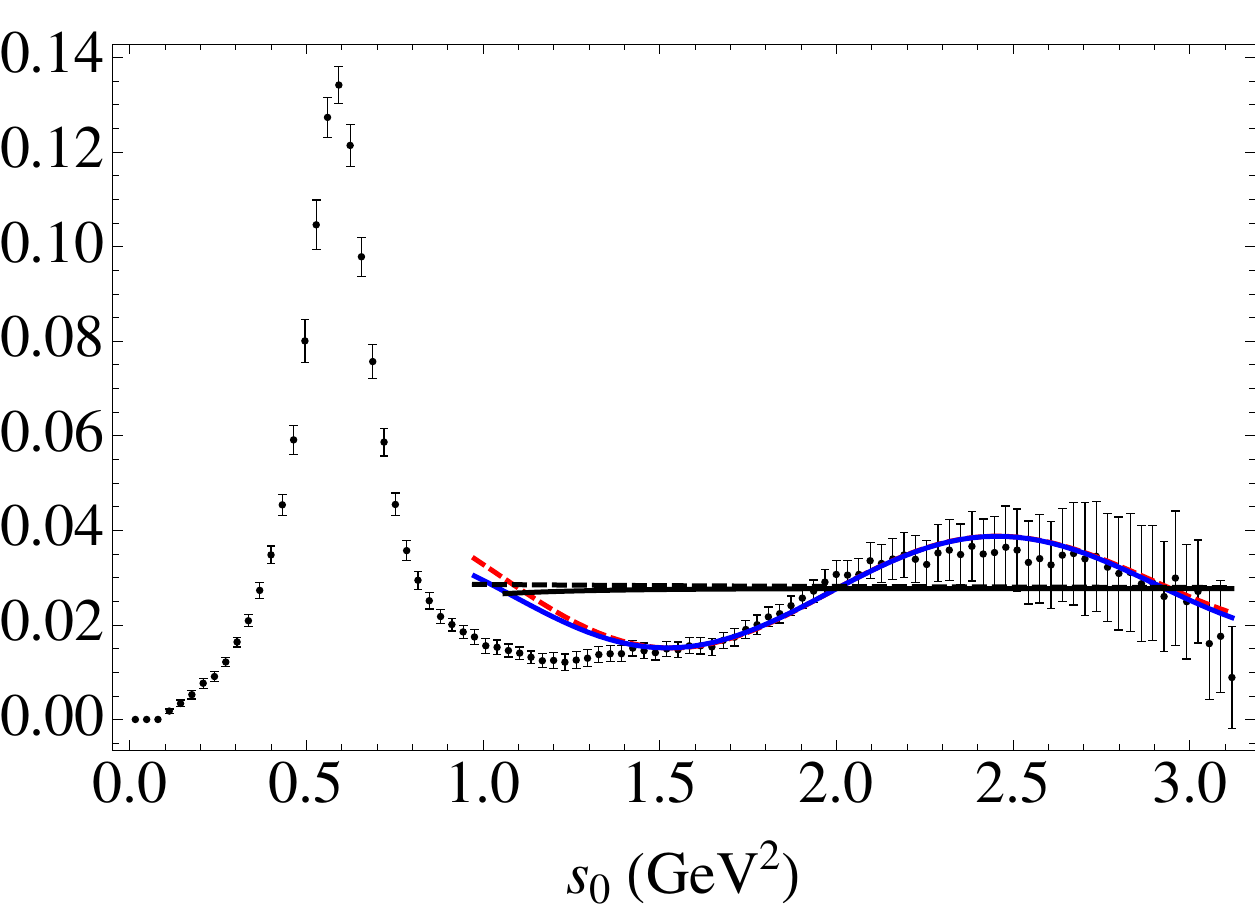}
\caption{Spectral function predicted by the fit of Fig.~\ref{VVw1}, compared with OPAL
data. CIPT fits are shown in red (dashed) and
FOPT in blue (solid).
The (much flatter) black curves represent the OPE parts.}
\label{VVspec}
\end{figure}

Figure~\ref{VVw1} shows the result of a fit to the vector channel with weight
$w=1$ and $s_{min}=1.5$~GeV$^2$, with all correlations between the data taken into account.  This fit leads to (FOPT = fixed-order perturbation theory, CIPT = contour-improved
perturbation theory)
\begin{eqnarray}
\label{alphas1}
\hspace{-0.6cm}\alpha_s(m_\tau^2)\!\!\!&=&\!\!\!0.307\pm 0.018\pm 0.004\pm 0.005\ ({\rm FOPT})\,,\\
&=&\!\!\!0.322\pm 0.025\pm 0.004\pm 0.005\ ({\rm CIPT})\ ,\nonumber\\
e^{-\delta}\!\!\!&=&\!\!\! 0.02\pm 0.01\ .\nonumber
\end{eqnarray}
The value of $\chi^2$ for this fit is 0.36 per degree of freedom.
The first error is the fit error, the second gives an indication of the stability
with respect to varying $s_{min}$, and the third estimates the effect of 
truncating perturbation theory.   {}From the result for $e^{-\delta}$ as well
as from Figs.~\ref{VVw1} and \ref{VVspec} it is clear that duality violations
have to be taken into account.   As mentioned above, and explained in detail
in Ref.~\cite{Betal1}, one may attempt to suppress duality violations by using
pinched weights, but at the price of not ignoring
terms in the OPE of dimension up to 16 if the weights of the standard
analysis are employed.   One should then worry, however, whether the OPE
converges to such high order for
$s_0\in[s_{min},m_\tau^2]$.

We have carried out a number of consistency checks:
\begin{itemize}
\item{} Fits with weights $w=1$, $1-(s/s_0)^2$ and $(1-s/s_0)^2(1+2s/s_0)$ and also including axial data lead to results completely consistent with
Eq.~(\ref{alphas1}).   These moments are preferred for their perturbative
behavior \cite{BBJ}.
\item{}  The first and second Weinberg sum rules, as well as the
sum rule for the electromagnetic pion mass difference are satisfied within
errors.
\item{} Our fits describe the non-strange hadronic tau-decay branching fraction
$R_{V+A,ud}(s_0)$ extremely well for $s_0$ above about 1.3~GeV$^2$.
\end{itemize}
The values found originally by OPAL are \cite{OPAL}:
\begin{eqnarray}
\label{alphasOPAL}
\alpha_s(m_\tau^2)&=&0.324\pm 0.014\quad ({\rm FOPT})\ ,\\
&=&0.348\pm 0.021\quad({\rm CIPT})\ .\nonumber
\end{eqnarray}
We find central values about $0.02$ lower on the same data,
and conclude that OPAL errors were underestimated.   In particular, errors due to
non-perturbative effects are at least as large as the difference between CIPT and FOPT.

\begin{figure}[t]
\includegraphics[width=.46\textwidth]{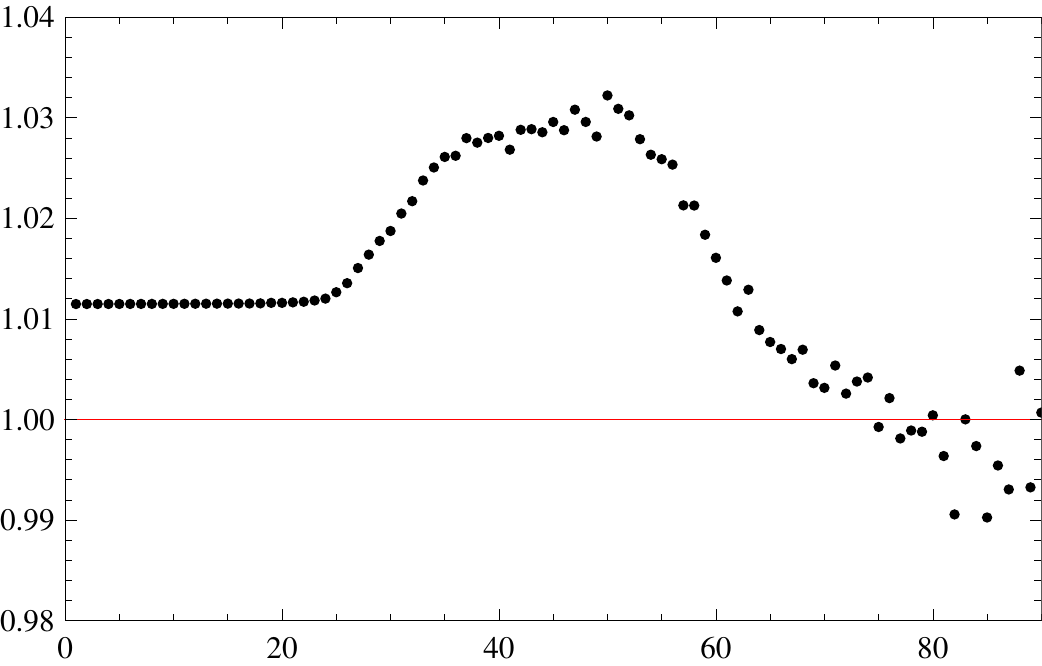}
\caption{Ratio of the vector-channel OPAL data using HFAG branching fractions and the original OPAL data using 1998 PDG branching fractions.
The horizontal axis labels the bins.}
\label{update}
\end{figure}

{\bf 4.}
The 1998 OPAL spectral functions were constructed as a sum over all
exclusive modes normalized with the 1998 PDG values of the
branching fractions.
In Ref.~\cite{Betal2}, we applied the analysis of Ref.~\cite{Betal1} to
rescaled OPAL data obtained by instead using current branching
fractions from HFAG \cite{HFAG}.  For the
vector channel, the rescaling factor is shown in Fig.~\ref{update}.

\begin{figure}[t]
\includegraphics[width=.80\textwidth]{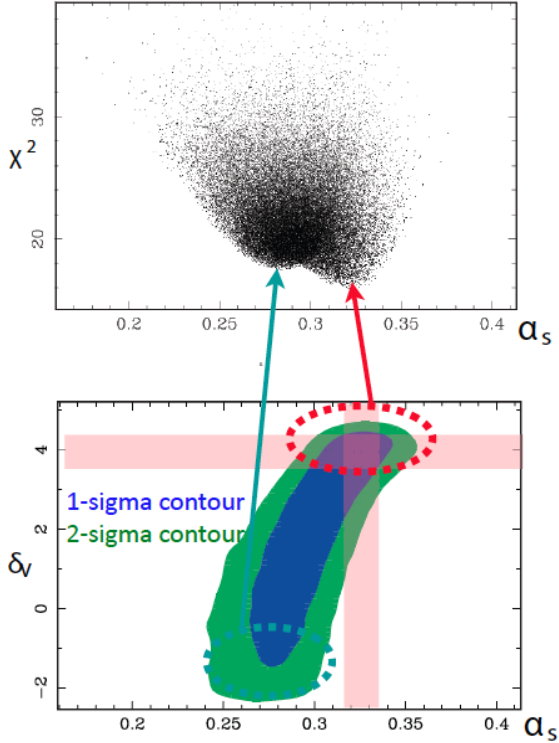}
\caption{Monte-Carlo analysis of the $\chi^2$ distribution for the fit of
Eq.~(\ref{sumrule}) with $w=1$ to the vector channel for the rescaled
OPAL data. $V$ stands for ``vector channel.''}
\label{diogo}
\end{figure}

In addition,
we also carried out a Markov-chain Monte Carlo analysis of the rescaled data
in order to explore the $\chi^2$ distribution as a function of the parameters
$\alpha_s(m^2_\tau)$ and those of Eq.~(\ref{ansatz}) in more detail.
This distribution forms a five-dimensional landscape, since it is a function of 5 parameters.
Cross sections of the full landscape are shown in Fig.~\ref{diogo}: in the top figure
we show the Monte-Carlo generated $\chi^2$ distribution projected onto the
$\alpha_s$--$\chi^2$ plane; in the bottom figure, we show the projection
onto the $\alpha_s$--$\delta_V$ plane.

We see that the $\chi^2$ distribution corresponds to a
rather complicated landscape.   The fit appears to allow for two different
minima, one with $\delta_V\approx 4$ and one with $\delta_V\approx -2$.
As can be seen in the lower figure, these two minima are not well separated,
and the value of $\chi^2$ per degree of freedom is reasonable for each
minimum.

While the absolute minimum is at $\delta_V\approx 4$, clearly we need
physical input in order to arrive at a value of $\alpha_s$ which is more
precise than a value in the range 0.27--0.35 or so that would follow from
the top figure.   The model developed in Ref.~\cite{CGP} favors 
\begin{equation}
\label{modeldelta}
\delta_V\approx -\log{\frac{F_\rho^2}{M_\rho^2}}\approx 4.2\ ,
\end{equation}
which leads us to choose the absolute minimum of the $\chi^2$ distribution
as the preferred value, and consider the other minimum unphysical.   With this choice, we find
\begin{eqnarray}
\label{alphas2}
\alpha_s(m_\tau^2)&=&0.325\pm 0.018\quad({\rm FOPT})\ ,\\
&=&0.347\pm 0.025\quad({\rm CIPT})\ .\nonumber
\end{eqnarray}
While the central values are virtually the same as those of Eq.~(\ref{alphasOPAL}), this is purely accidental.
Moreover, as before, the errors are larger, because of non-perturbative
effects.   Indeed, if we define $\delta^{\rm NP}$ by
\begin{equation}
\label{NP}
R_{V+A,ud}=N_c|V_{ud}|^2\left(1+\delta^{pert.th.}+\delta^{\rm NP}\right)\ ,
\end{equation}
we find that 
\begin{eqnarray}
\label{NPvalues}
\delta^{\rm NP}&=&-0.004\pm 0.012\quad(\mbox{FOPT})\ ,\\
&=&-0.002\pm 0.012\quad(\mbox{CIPT})\ .\nonumber
\end{eqnarray}
This is to be compared with $\delta^{\rm NP}=-0.0059
\pm 0.0014$ from the standard analysis \cite{Detal08,Pich}.   Our error on
$\delta^{\rm NP}$ is an
order of magnitude larger, because non-perturbative effects
have been treated systematically in our analysis.

{\bf 5.} 
We presented a new analysis of hadronic tau decays, yielding a new 
value for the strong coupling, $\alpha_s$, at the tau mass, with a larger
error than found in previous analyses.   Earlier values were all based on the
standard analysis summarized in Sec.~3, and, moreover, on the incomplete
ALEPH data.   Therefore, we believe that our values should be taken as
superseding all earlier values for $\alpha_s$ from hadronic tau decays.

As we saw in Sec.~4, fits to OPAL data are at the very edge of what is
statistically possible.   This is not a flaw of the analysis, but appears to be
the best one can expect based on the OPAL data.   In Ref.~\cite{Betal2},
we investigated what would happen with the $\chi^2$ distribution shown in
Fig.~\ref{diogo}  if the errors are reduced by a factor 
2 or 3.   We found that the degeneracy shown in Fig.~\ref{diogo} disappears.   Therefore, we
expect that a much more stringent test of the sum-rule analysis of
hadronic tau decays would be possible if inclusive spectral functions
were made available from the BaBar or Belle data.

{\bf Acknowledgements}
We would like to thank M.~Beneke, C.~Bernard, A.~H\"ocker,
M.~Martinez, and R.~Miquel for discussions, and
S.~Banerjee and S.~Menke for help with understanding the
HFAG analysis and OPAL  data, respectively.
DB is
supported by the Alexander von Humboldt Foundation, and
MG  in part by the US Dept. of Energy, and in part by the Spanish Ministerio de Educaci\'on, Cultura y Deporte, under program SAB2011-0074.
MJ and SP are supported by CICYTFEDER-FPA2008-01430, FPA2011-25948, SGR2009-894,
the Spanish Consolider-Ingenio 2010 Program
CPAN (CSD2007-00042). 
AM was supported in part by NASA through Chandra award No.~AR0-11016A.
KM is supported by the Natural Sciences and
Engineering Research Council of Canada.

\end{document}